\documentclass{aa}

\usepackage{graphicx}
\usepackage{txfonts}
\usepackage{natbib}     
\usepackage{xcolor} 

\bibpunct{(}{)}{;}{a}{}{,} 

\begin{document}

   \title{The centre-to-limb variations of solar Fraunhofer lines imprinted upon lunar eclipse spectra}

   \subtitle{Implications for exoplanet transit observations}

   \author{F. Yan
          \inst{1,2,3}
          \and
          R. A. E. Fosbury
          \inst{3}
          \and
          M. G. Petr-Gotzens
          \inst{3}
          \and
          G. Zhao
          \inst{1}
          \and
          E. Pall\'e
          \inst{4,5}
                }

   \institute{Key Laboratory of Optical Astronomy, National Astronomical Observatories, Chinese Academy of Sciences, 20A Datun Road, Chaoyang District, 100012 Beijing, China\\
\email{feiy@nao.cas.cn, gzhao@nao.cas.cn}
         \and
                University of Chinese Academy of Sciences, 19A Yuquan Road, Shijingshan District, 100049 Beijing, China
         \and
                        European Southern Observatory, Karl-Schwarzschild-Str. 2, 85748 Garching bei M\"unchen, Germany
                \and
                        Instituto de Astrof\'isica de Canarias, C/ v\'ia L\'actea, s/n, 38205 La Laguna, Tenerife, Spain
        \and                    
                        Dpto. de Astrof\'isica, Universidad de La Laguna, 38206 La Laguna, Tenerife, Spain
\\      
         }
        \date{Received -----; accepted -----}


  \abstract
  {The atmospheres of exoplanets are commonly studied by observing the transit of the planet passing in front of its parent star. The obscuration of part of the stellar disk during a transit will reveal aspects of its surface structure resulting from general centre-to-limb variations (CLVs). These become apparent when forming the ratio between the  stellar light in and out of transit. These phenomena can be seen particularly clearly during the progress of a penumbral lunar eclipse, where the Earth transits the solar disk and masks different regions of the solar disk as the eclipse progresses. When inferring the properties of the planetary atmosphere, it is essential that this effect originating at the star is properly accounted for. 
  Using the data observed from the 2014-April-15 lunar eclipse with the ESPaDOnS spectrograph mounted on the Canada France Hawaii Telescope (CFHT), we have obtained for the first time a time sequence of the penumbral spectra. These penumbral spectra enable us to study the centre-to-limb variations of solar Fraunhofer lines when the Earth is transiting Sun. 
The \ion{Na}{i} and \ion{Ca}{ii} absorption features reported from previous lunar eclipse observations are demonstrated to be CLV features, which dominate the corresponding line profiles and mask possible planetary signal.
  Detecting atmospheric species in exoplanets via transit spectroscopy must account for the CLV effect.
}


   \maketitle

%

\section{Introduction}
With the discovery of almost 2000 exoplanets in the last two decades, the characterisation of their atmospheres has become a new and rapidly-evolving field.
Different  atomic, ionic, and molecular species have already been detected in exoplanet atmospheres. Using Hubble Space Telescope data,
\cite{Charbonneau2002} detected sodium in the atmosphere of \object{HD 209458b} for the first time by comparing the \ion{Na}{i} doublet absorption lines in and out of transit. \cite{Snellen2008} then confirmed the \ion{Na}{i} absorption in \object{HD 209458b} using ground-based telescope data. Sodium has also been detected in several other giant exoplanets such as \object{HD 189733b} \citep{Redfield2008} and \object{WASP-17b} \citep{Wood2011}.
Potassium was detected in \object{HD 80606b} \citep{Colon2012} and \object{XO-2b} \citep{Sing2011}, while \cite{Fossati2010} reported the detection of \ion{Mg}{ii} in \object{WASP-12b}. All of these used transmssion spectroscopy during transits.
Molecular features have also been detected in exoplanet atmospheres, such as $\mathrm{H_2O}$ \citep{Grillmair2008}, $\mathrm{CO}$ \citep{Brogi2012}, $\mathrm{CO_2}$ \citep{Swain2009}, and $\mathrm{CH_4}$ \citep{Swain2008}.

In recent years, nearly 100 Earth-sized or smaller exoplanets have been discovered, one of them being in the habitable zone of its host star \citep{Quintana2014}. 
Although characterising the atmospheres of these terrestrial exoplanets is not within the reach of current instrumentation, the detection of some their atomic and molecular components is very likely to become possible with the next generation of large ground- and space-based telescopes \citep{Hedelt2013}. 
The Earth itself can be used as a benchmark for the future detection of Earth-like exoplanets, for example, using lunar eclipses to obtain the tangential long-path transmission spectrum of the Earth's atmosphere.

Several lunar eclipse observations have been made with the aim of obtaining the transmission spectrum of the Earth's atmosphere. Some of these have reported the anomalous behaviour of certain atomic absorption features. 
\citet[hereafter P09]{Palle2009} observed the lunar eclipse of 16 August 2008, and obtained transmission spectra from the umbral lunar eclipse. In P09’s spectra, the \ion{Ca}{ii} absorption features are detected and weak \ion{Na}{i} absorptions also appear. \citet[hereafter V10]{Vidal2010} observed the same lunar eclipse as P09, but they retrieved the transmission spectrum from the penumbral rather than the umbral eclipse. In V10, the authors detected relatively strong \ion{Na}{i} D lines absorption but no \ion{Ca}{ii} absorption.  
\citet[hereafter A14]{Arnold2014} observed the penumbral lunar eclipse in December 2010 and their results are similar to those of V10, i.e. \ion{Na}{i} absorption is detected while \ion{Ca}{ii} is not.
\citet{Yan2014} observed the lunar eclipse of December 2011 and neither the \ion{Na}{i} nor the \ion{Ca}{ii} absorption features are detected in the transmission spectrum obtained from the umbral eclipse. 

The interpretation of these discrepancies is not straightforward since the Earth’s transmission spectra from these observations are retrieved using different methods. Although there are \ion{Na}{i}, \ion{Ca}{i,} and \ion{Ca}{ii} layers in the Earth’s ionosphere, according to our research, the \ion{Na}{i} or \ion{Ca}{ii} absorption features in the observed Earth's transmission spectra are probably due to the centre-to-limb variation (CLV) of the solar lines rather than the absorptions in the Earth’s atmosphere. 

The strong solar Fraunhofer lines have prominent variations in both line intensity and profile from centre to limb cross the solar disk.
\cite{Athay1972} used the \ion{Fe}{i} line's CLV spectrum to generally interpret the CLV effect. For most strong solar Fraunhofer lines, the normalised spectral line from the centre disk is deeper than that from the limb. The observed CLV features are used to understand detailed solar physics and to guide the modelling work. For example, \cite{Allende2004} observed the CLV of solar lines and used this to test non-Local Thermodynamic Equilibrium line formation calculations and \cite{Koesterke2008} used the CLV data to test 3D solar hydrodynamic simulations. 

Since a lunar eclipse shares similarities with an exoplanet transit, this CLV effect can also be present in transit spectra. The corresponding features will be considerably weaker since the planet only blocks a small fraction of the stellar surface during transit. However, the CLV effect can still have an influence on the interpretation of
atomic line detections in exoplanet atmospheres and should be properly treated. 

For the detection of sodium in HD 209458b, \cite{Charbonneau2002} considered the CLV in the \ion{Na}{i} D lines and concluded that the contribution was small for their observation. Redfield et al. (2008) modelled the CLV effect on the  \ion{Na}{i} D lines in HD 189733b (the CLV is referred to as differential limb darkening in their paper), and the model shows that this contribution is much smaller than the observed \ion{Na}{i} absorption. However, for stellar lines that have significant CLV effects or if the observation is performed at high spectral resolution, this effect could become important.

We observed the lunar eclipse in April 2014 and obtained a set of spectra at different stages of the eclipse. The changes in the solar line profiles caused by the CLV can  clearly be seen in our data.
In Section 2 , we use our lunar eclipse data to demonstrate this effect and compare it with the solar spectral Atlas.
In Section 3, the \ion{Na}{i} and \ion{Ca}{ii} features observed in previous lunar eclipse observations are compared with the spectral features caused by CLV.
In Section 4, we further discuss the CLV effect on exoplanet transit spectroscopy.
The telluric sodium absorption and the Raman scattering in lunar eclipse spectra are also discussed in Section 4.

   \begin{figure}
   \centering
   \includegraphics[width=0.50\textwidth]{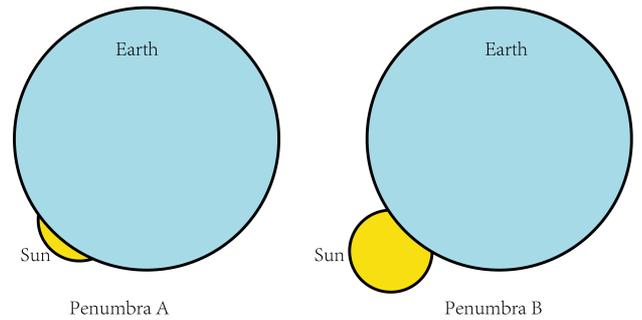}
      \caption{A schematic of a penumbral eclipse as seen from the Moon. Here Penumbra A and Penumbra B represent the views from two different penumbral locations (indicated in Fig.~\ref{NASA-AB}). }
         \label{PMview}
   \end{figure}
%


\section{The effect of CLV on lunar eclipse observations}
A penumbral lunar eclipse happens when the Moon enters the Earth's penumbral shadow. An observer on the Moon in the penumbra would see the Earth blocking different parts of the solar disk (as shown in Fig.~\ref{PMview}). Thus the penumbral spectrum comprises the integrated spectrum of the Sun, which is partly eclipsed by the Earth. 
When observing different locations on the Moon in the Earth's penumbra, the observed spectra are integrated from different parts of the solar disk. Since the solar lines vary in profile from the centre to limb, the line profiles in the penumbral spectra vary as the eclipse proceeds.
For the spectrum observed from the un-eclipsed Moon (which is called the bright
Moon), it is the integration of the entire solar disk.
The umbral spectrum results from the integrated light of the solar disk that is refracted by the Earth's atmosphere into the umbral shadow \citep{Garcia2012}. The line shape in the umbral spectrum varies, depending on which part of the solar disk has been refracted.
However, the line shape difference between the umbral spectrum and the bright Moon spectrum is relatively small, and  the penumbral spectrum exhibits significant differences (see Fig.~\ref{Ha-change} for details).

\subsection{The April 2014 lunar eclipse observations}
\begin{figure}
   \centering
   \includegraphics[width=0.550\textwidth]{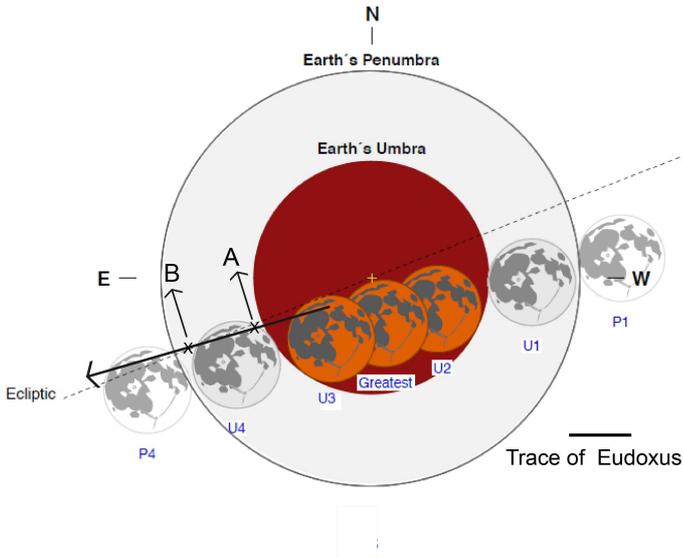}
      \caption{The trajectory of the observed Eudoxus crater for the 15 April 2014 lunar eclipse. The arrow on the trajectory indicates the moving direction of the crater. Here we choose two points on the trajectory (A and B) to indicate different eclipsing stages. Point A and B correspond to the Penumbra A and Penumbra B views sketched in Fig.~\ref{PMview}. 
      The figure is reproduced from the NASA eclipse website (http://eclipse.gsfc.nasa.gov/lunar.html).}
         \label{NASA-AB}
   \end{figure}

We used the fiber-fed ESPaDOnS spectrograph mounted on the Canada France Hawaii Telescope (CFHT)  for our observations. The spectrograph covers a wavelength range of 370--1050 nm in a single exposure.
The `object only' mode was used to achieve high spectral resolving power (average resolution: $\lambda / \Delta \lambda$ $\sim$ 81,000). The observation began when the Moon was in the Earth's umbra and lasted until the Moon was fully out of the Earth's shadow. 
We used non-sidereal tracking  to locate the fiber at the Eudoxus crater during the entire observation. 
The Eudoxus crater is chosen because it enables us to access different phases of the umbral eclipse as its trace is close to the umbral center.
Fig.~\ref{NASA-AB} shows the trace of Eudoxus with respect to the Earth's shadow.
We obtained a sequence of spectra from the umbra, penumbra, and bright Moon,  and we performed the standard data reduction procedure  using the CFHT Upena pipeline \footnote{http://www.cfht.hawaii.edu/Instruments/Upena/}. 

\begin{figure}
   \centering
   \includegraphics[width=0.50\textwidth]{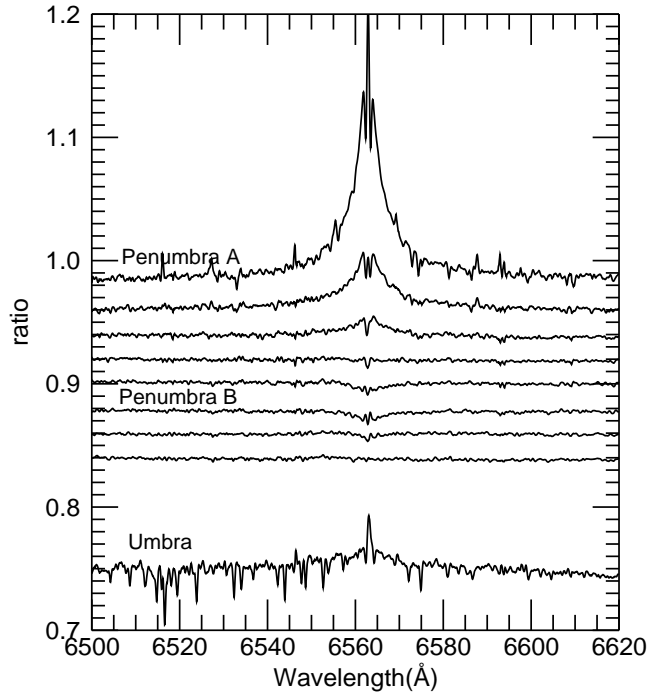}
      \caption{The change of H$\alpha$ line shape in different penumbra spectra. The spectra are the ratios of the normalised penumbral to bright Moon spectra. The ratios are shifted sequentially by 0.02 for clarity.
Eight penumbral spectra which were taken every ten minutes are chosen to represent different penumbral locations.
The upper spectrum corresponds to the penumbral position close to the umbra (Penumbra A in Fig.~\ref{NASA-AB}), while the lower spectrum corresponds to the position close to the bright Moon (Penumbra B in Fig.~\ref{NASA-AB}). 
The normalised ratio of the umbral to the bright Moon spectrum is also shown at the bottom of the figure for comparison. This umbral spectrum is the sum of seven individual exposures that were taken at positions close to the edge of the umbra (i.~e. close to Penumbra A). The total observed time for this umbral spectrum is about 20 minutes. The umbral spectrum has a relatively low signal-to-noise ratio and strong telluric absorption lines.
The radial velocity differences between these spectra have been corrected.
}
         \label{Ha-change}
   \end{figure}

The penumbral spectral sequence consists of about 60 spectra from different penumbral locations, enabling us to study the line shape change. We label the location close to the umbra as Penumbra A and the location close to the bright Moon as Penumbra B (Fig.\ref{PMview}~ and Fig.~\ref{NASA-AB}). 
Fig.~\ref{Ha-change} shows the changes of the H$\alpha$ line profile. The spectra in this figure are the ratios of the penumbral spectra to a bright Moon spectrum. 
The bottom of this figure shows the ratio of a typical umbral to the bright Moon spectrum for comparison.

At the beginning of our penumbral eclipse observation (corresponding to Penumbra A), the line shape is generally shallower than that in the bright Moon spectrum, which results in the `emission' feature in Fig.~\ref{Ha-change}. As the eclipse proceeds, the H$\alpha$ line becomes deeper. When the observed location is close to Penumbra B, the H$\alpha$ line becomes the deepest, even deeper than in the bright Moon spectrum.
As this change of the H$\alpha$ line shape is correlated with the eclipse geometry, 
we conclude that it results from the CLV effect.

Since previous studies of lunar eclipse observations use the penumbral spectrum obtained at a position close to the umbra (i.e. Penumbra A), we will only discuss this type of penumbral spectrum below.

%

%

   \begin{figure*}
   \centering
   \includegraphics[width=0.98\textwidth]{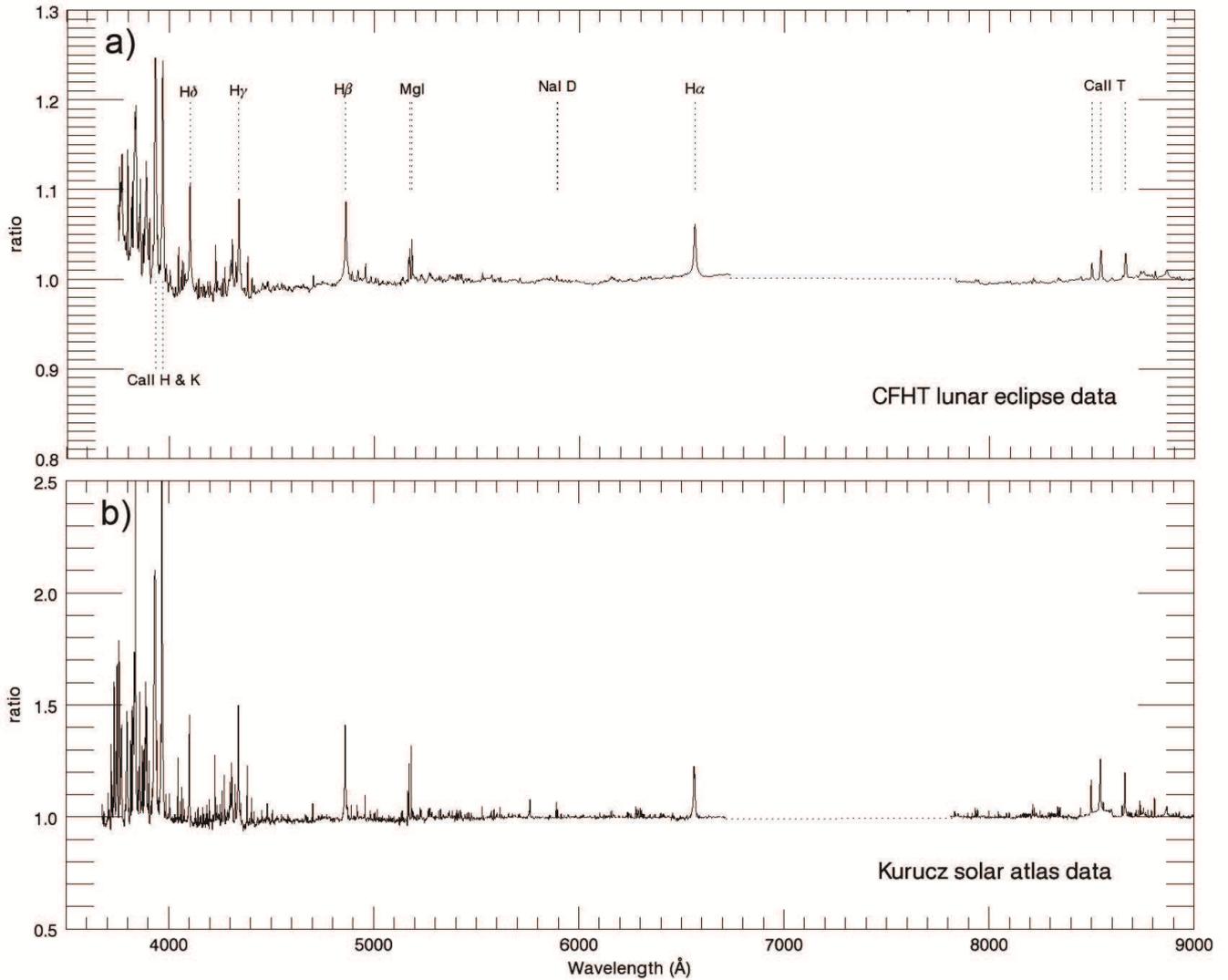}
\caption{(a)The ratio of a penumbral spectrum to a bright Moon spectrum observed on April 15 2014 by CFHT. The solar lines features remain in the ratio spectrum. The data is binned to a low resolution and the RV difference between the two spectra has been corrected. The dashed line region is where the telluric $\mathrm{H_2O}$ and $\mathrm{O_2}$ lines are prominent. They are not shown here as we concentrate on the solar spectral feature.
(b)The ratio of the limb solar spectrum to the centre solar spectrum. The data are from the Kitt Peak Solar Atlas and binned to a low resolution.
}

         \label{overall}
   \end{figure*}

\subsection{CLV features in the spectrum}
Fig.~\ref{overall}a displays the ratio of an observed penumbral to a bright Moon spectrum. Here the penumbral spectrum was taken when the observed location was close to the umbra, which means the penumbral spectrum is mainly from the limb part of the solar disk (corresponding to Penumbra A in Fig.~\ref{PMview}). 
In contrast, the bright Moon spectrum is from the whole solar disk and so it is the mix of limb and centre solar spectra. This spectral ratio can therefore be qualitatively regarded as the limb to centre spectrum plus a constant.
It can be seen in Fig.~\ref{overall}a that the solar line features clearly show up especially towards the blue part of the spectrum. Strong absorption lines such as the \ion{Ca}{ii} T (\ion{Ca}{ii} near-infrared triplet), \ion{Ca}{ii} H \& K, H$\alpha$, H$\beta,$ and the \ion{Mg}{i} b lines appear prominently while the \ion{Na}{i} D lines features are relatively weak. 

To better demonstrate the CLV effect in a lunar eclipse, the solar spectrum from the Kitt Peak Solar Atlas \citep{Brault1972, Kurucz2005} is used as a comparison. 
The Atlas comprises two sets of spectra: one observed at the centre of the solar disk with $\mu=1.0$ (where $\mu=\mathrm{cos(}\theta)$, $\theta$ is the angle between the normal to the solar surface and the line of sight) and the other at a limb position with $\mu=0.2$. All the spectra were taken when the Sun was quiet.
The ratio of the centre to limb Solar Atlas is shown in Fig.~\ref{overall}b and one can see that the overall features are very similar to those in Fig.~\ref{overall}a.

These general features show that the strong Fraunhofer lines are deeper in the centre than in the limb spectrum, which means the mean limb-darkening within the spectral line is less pronounced than in the adjacent continuum \citep{Athay1972book}.
The continuum limb darkening is more prominent at blue/UV wavelengths because of the amplification of the temperature induced intensity variation in the Wien side (short wavelength region) of the blackbody curve.
This also makes the mean CLV features become more significant for the lines located at the blue/UV wavelengths.

In addition to the general CLV features described above, the detailed profile has a complicated structure, which differs from line to line.
For example, the line core becomes broader in the limb spectrum than in the centre spectrum, which makes the line core CLV feature differ from that of the line wing. This can be seen for example in the \ion{Ca}{ii} T and \ion{Na}{i} D lines shown in Fig.~\ref{Ca8662} and \ref{Na-compare}. The line core CLV is due to the Doppler broadening, which is stronger in the limb than in the centre spectrum. This differential Doppler broadening may be interpreted either as an increased Doppler velocity with height or as an anisotropy in the Doppler velocity field where the horizontal velocity exceeds the vertical velocity \citep{Athay1972book}. 
Other solar phenomena such as the convective motion of the granules \citep{Dravins1982} also contribute to the detailed CLV features, but these are beyond the scope of this paper.

\section{\ion{Ca}{ii} and \ion{Na}{i} features in lunar eclipse observations}
\begin{figure}
   \centering
   \includegraphics[width=0.5\textwidth]{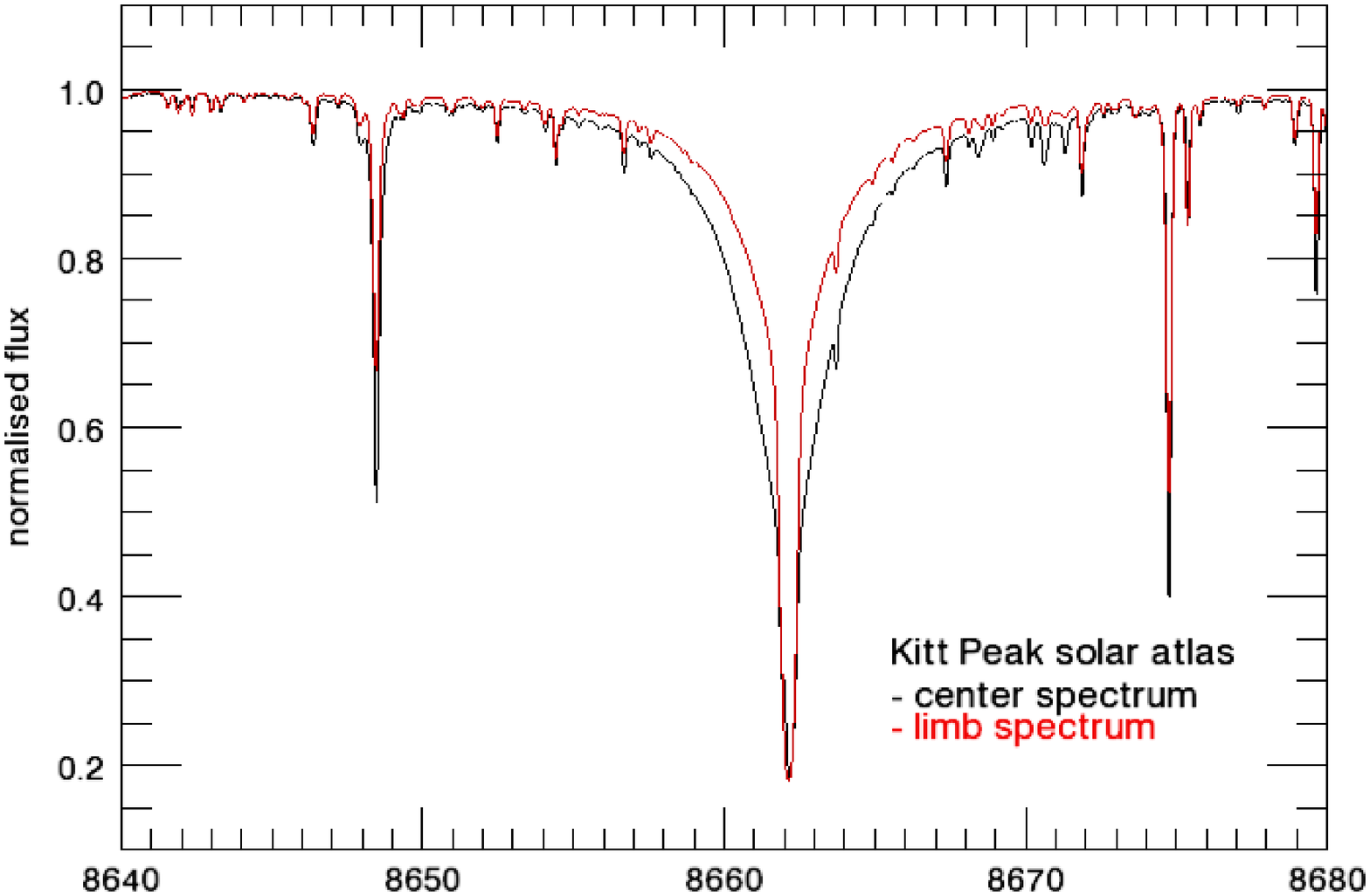}
   \includegraphics[width=0.5\textwidth]{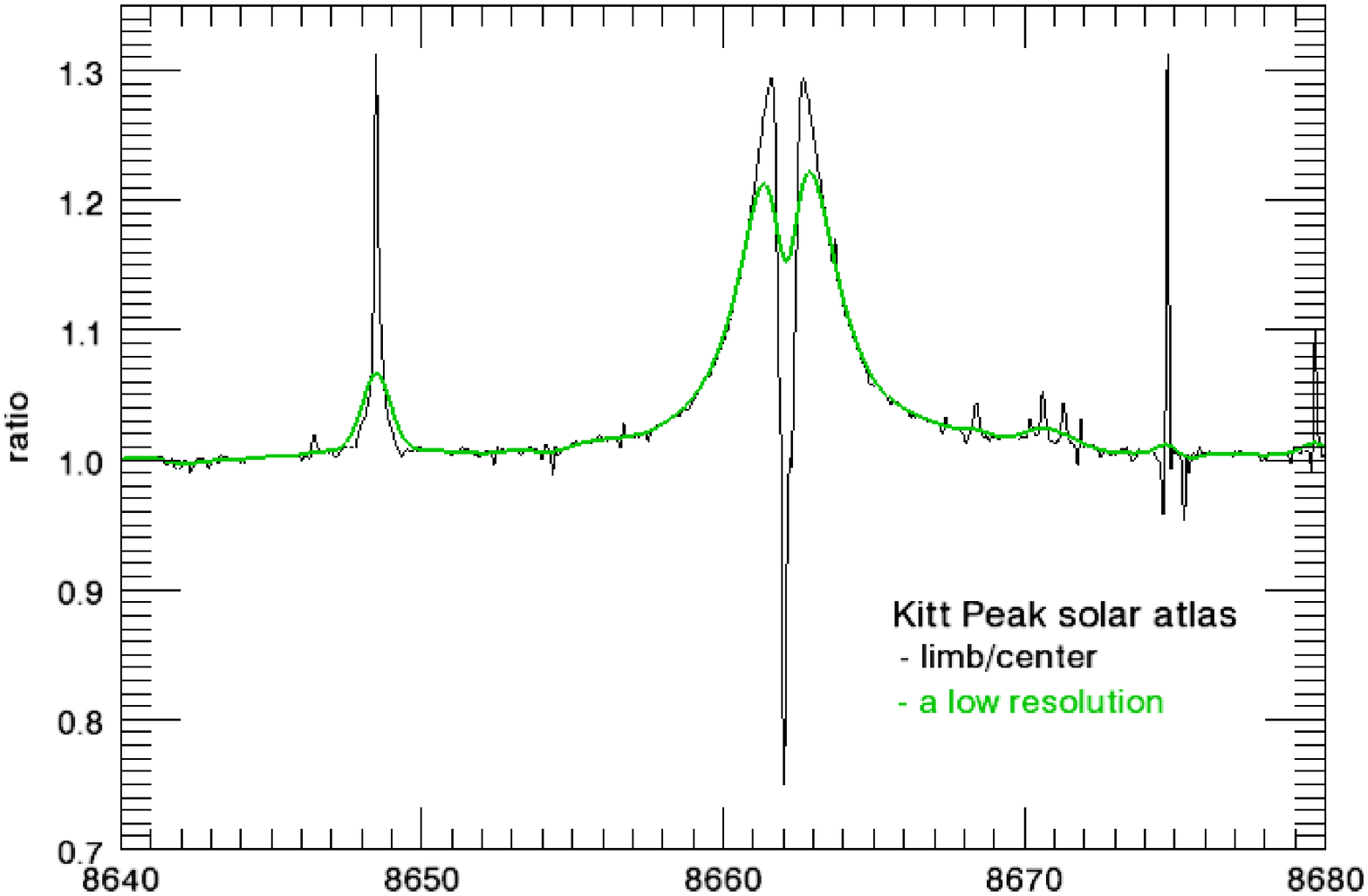}
   \includegraphics[width=0.5\textwidth]{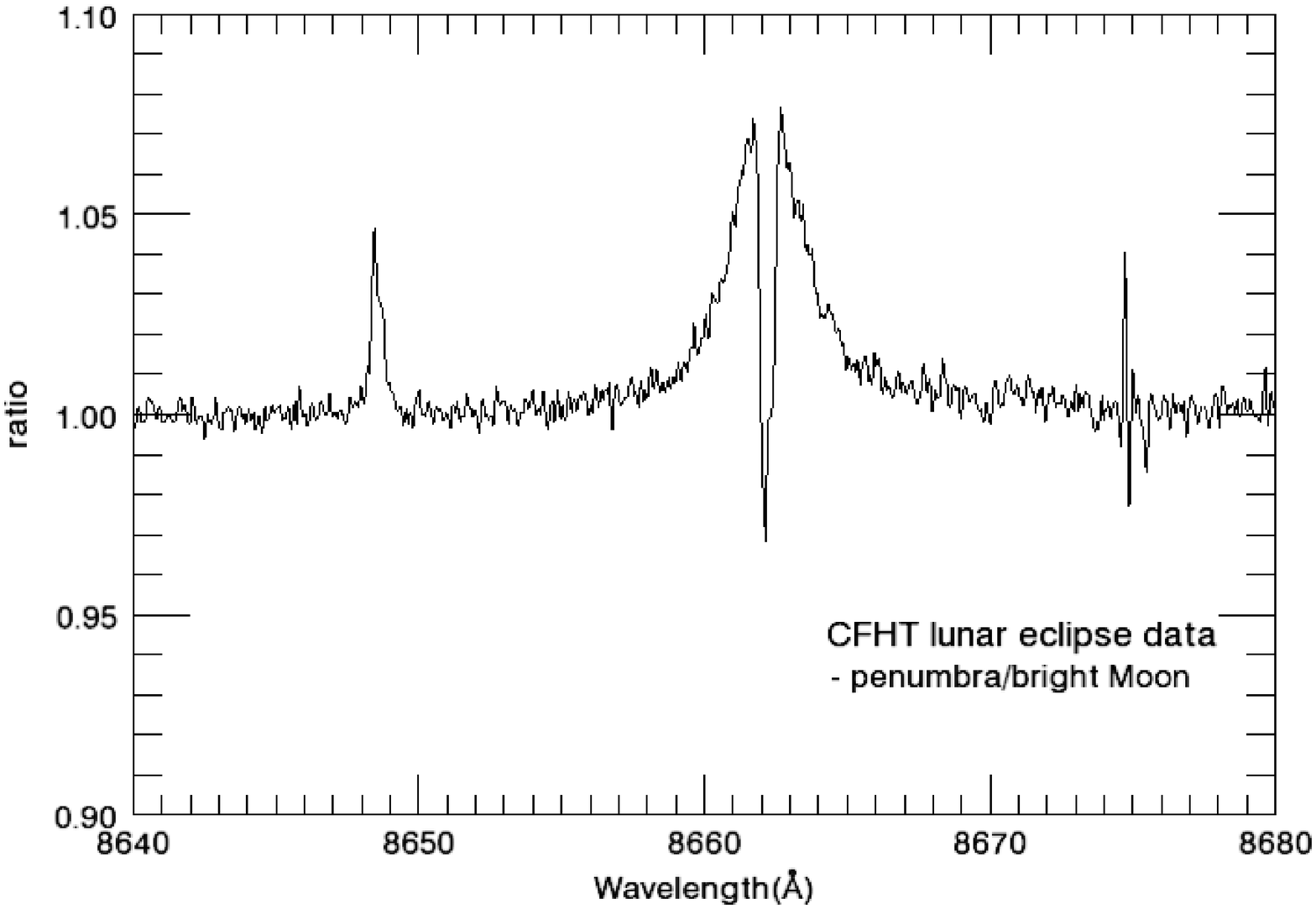}
\caption{The detailed line shape of \ion{Ca}{ii} 8662$\,\AA$ line. Top panel: the limb and centre spectra from the Kitt Peak Solar Atlas. 
Middle panel: the ratio of the limb to centre Kitt Peak spectrum. The green line is same data convolved to 1 $\,\AA$ resolution.
Bottom panel: the ratio of the penumbra to bright Moon spectrum from CFHT lunar eclipse data (the same data set as shown in Fig \ref{overall}a).
}
         \label{Ca8662}
   \end{figure}

\begin{figure}
   \centering
   \includegraphics[width=0.5\textwidth]{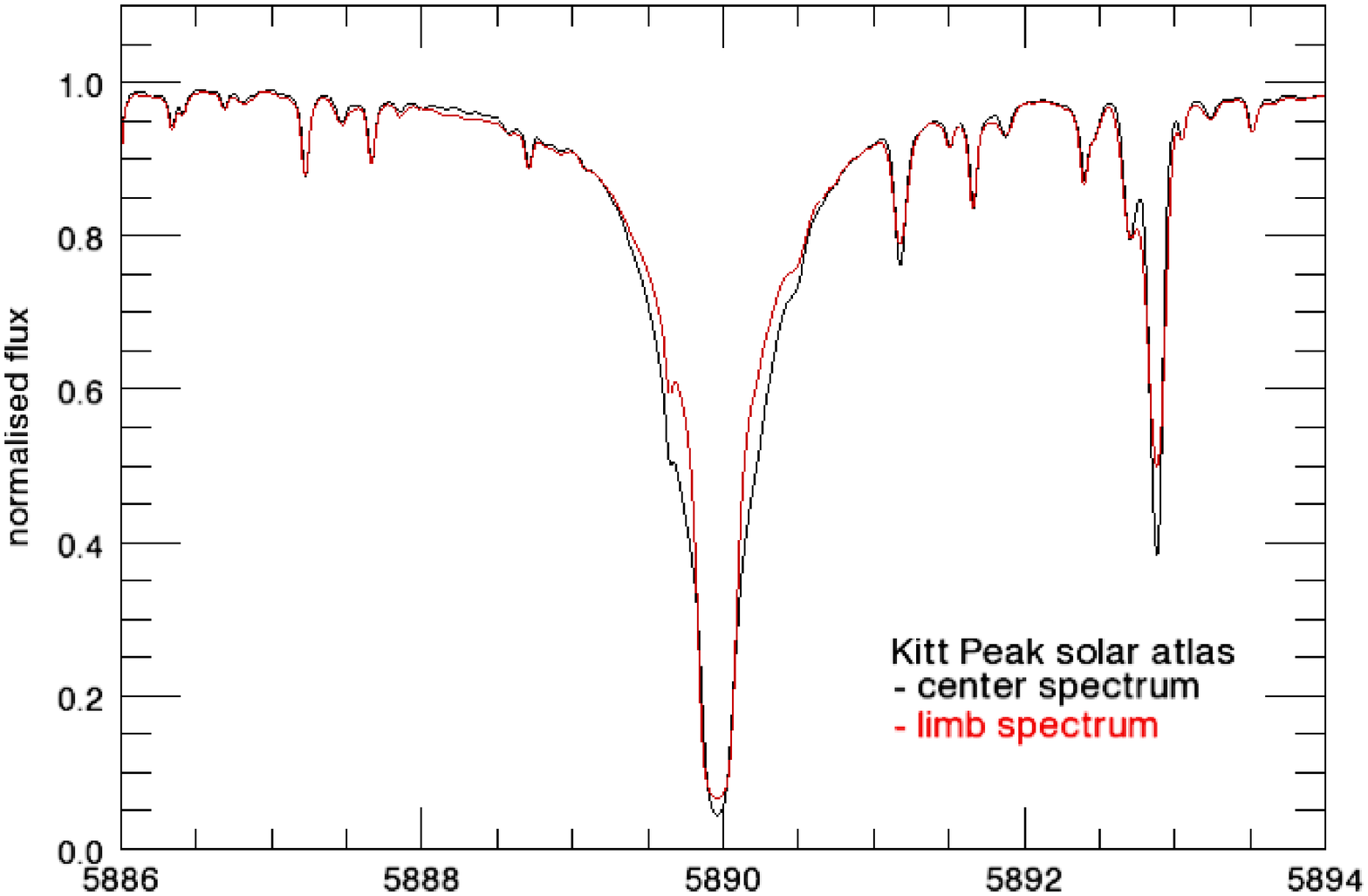}
   \includegraphics[width=0.5\textwidth]{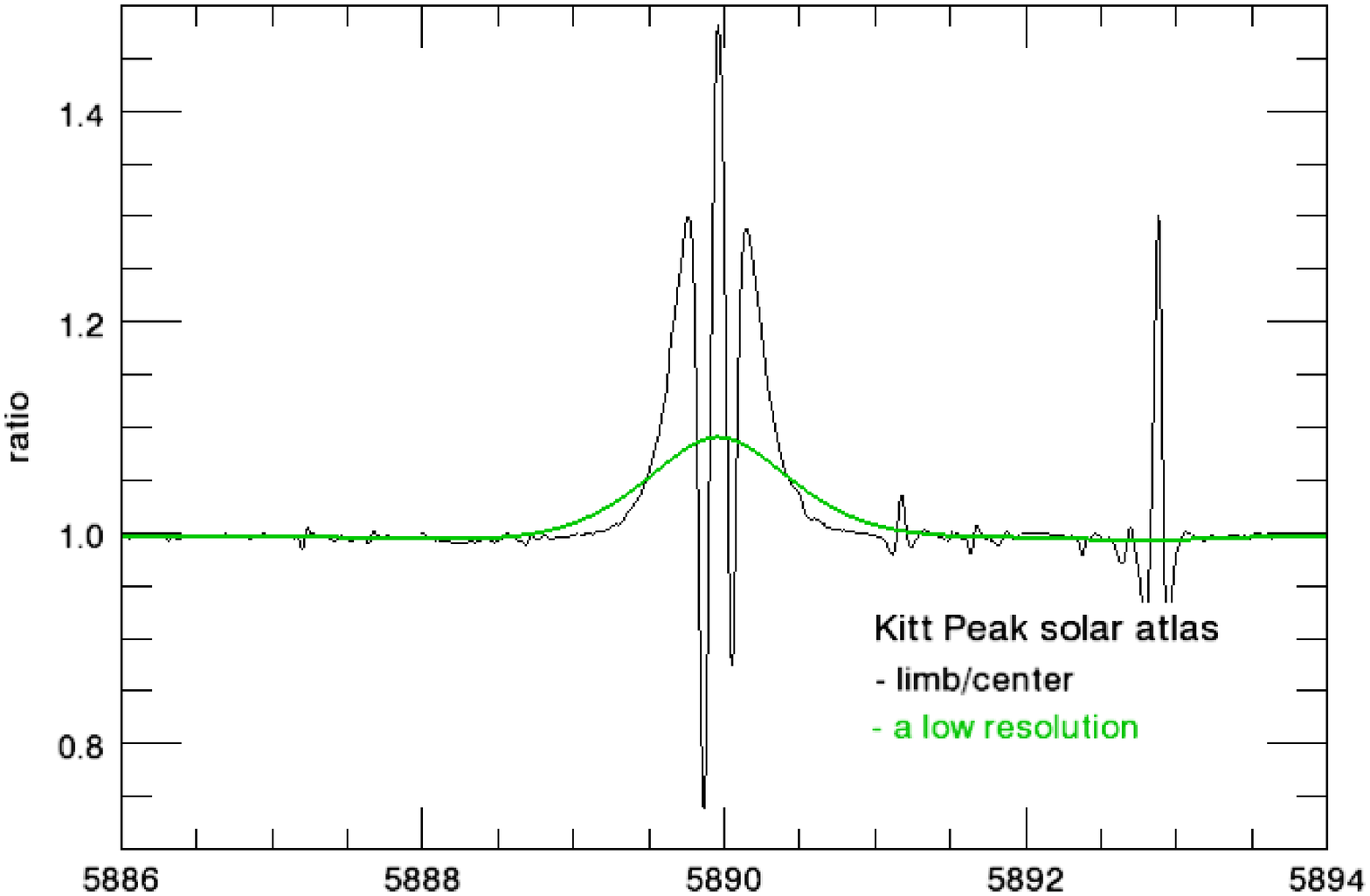}
   \includegraphics[width=0.5\textwidth]{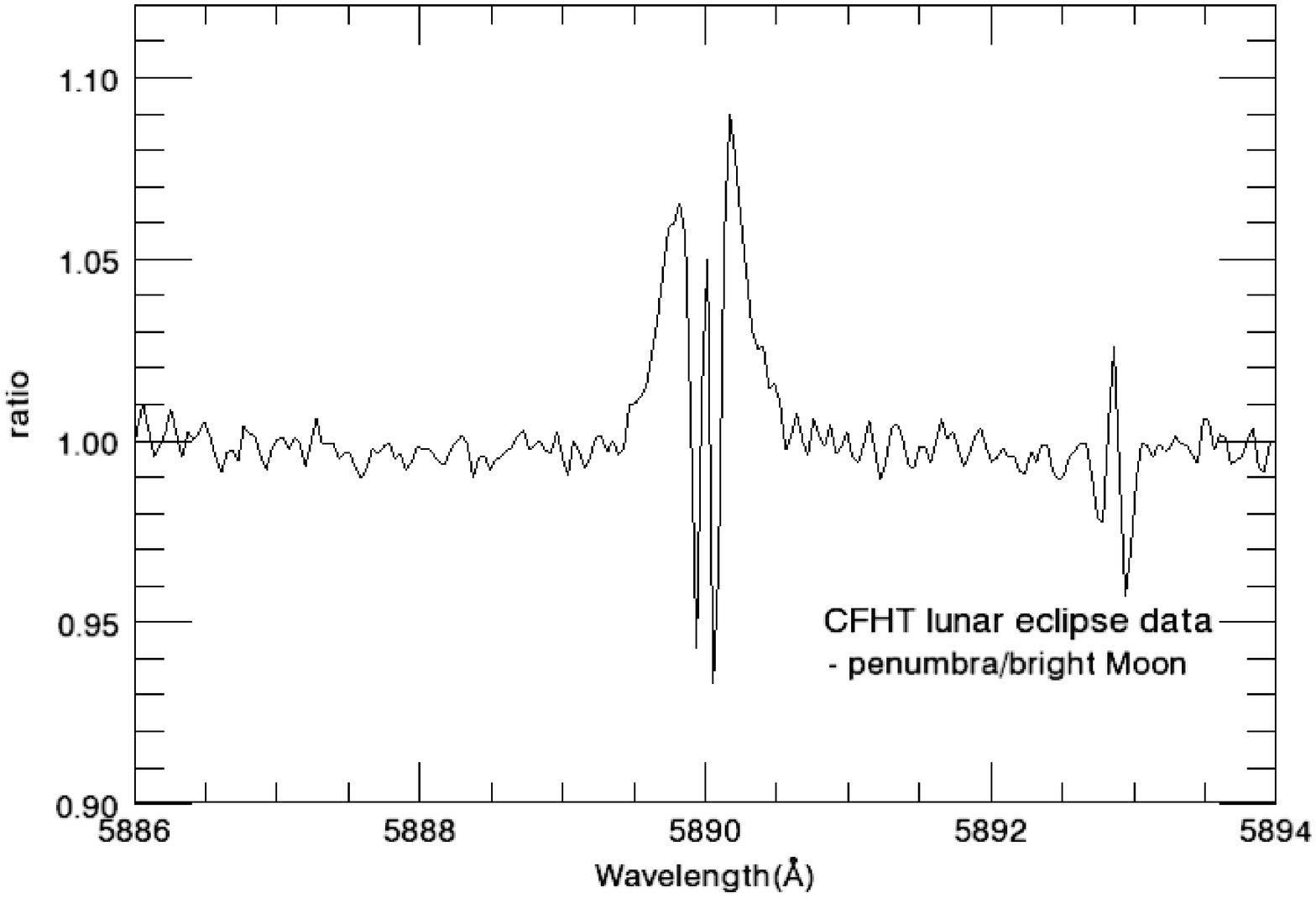}
\caption{The same as Fig.~\ref{Ca8662} but for \ion{Na}{i} $\mathrm{D_2}$ line at 5890$\,\AA$ .
}
         \label{Na-compare}
   \end{figure}

Since ionised calcium and neutral sodium are present in the Earth's ionosphere and previous studies of lunar eclipse observations claimed the detection of their absorption lines, it is important to understand how the solar CLV affects their detections in the Earth's transmission spectrum.
For the CLV of the \ion{Ca}{ii} Fraunhofer lines (e.g. the \ion{Ca}{ii} T and \ion{Ca}{ii} H\&K), we choose the \ion{Ca}{ii} 8662$\,\AA$ line as a demonstration. Fig.~\ref{Ca8662} shows the 8662$\,\AA$ line from both the Kitt Peak Solar Atlas
and the CFHT lunar eclipse data. 
The ratio of the solar limb to centre spectrum (middle panel in Fig.~\ref{Ca8662}) and the ratio of penumbra to bright Moon spectrum (bottom panel) are quite similar.  Fig.~\ref{Na-compare} shows the \ion{Na}{i} $\mathrm{D_2}$ line at 5889$\,\AA$. The CLV feature of the \ion{Na}{i} $\mathrm{D_2}$ line is similar to, but much weaker than, the \ion{Ca}{ii} T lines. 
In the middle panel of this figure, the one $\,\AA$ resolution spectrum (shown as the green line) smears the feature, indicating how the instrumental resolution can affect the observed strength of the CLV feature.

\subsection{\ion{Ca}{ii} and \ion{Na}{i} features in previous low spectral resolution observation}
\citet{Palle2009} observed the 16 August 2008 lunar eclipse at a relatively low resolution (6.8 $\AA$ in the optical wavelength range). They regarded the penumbral spectrum as the reference spectrum and obtained the Earth's transmission spectrum by using the ratio of the umbral to the penumbral spectrum to cancel out the solar spectral features. However, because of the CLV, the solar absorption line profiles are different in the umbral and penumbral spectra. The solar line shapes in the umbral spectrum are similar to the bright Moon spectrum while the line shapes in the penumbral spectrum have a significant CLV effect. Thus the \ion{Ca}{ii} T and \ion{Ca}{ii} H\&K lines, as well as the weak \ion{Na}{i} D lines absorption features in P09's transmission spectrum, are most likely to be the result of the CLV effect.

\begin{figure*}
   \centering
   \includegraphics[width=0.98\textwidth]{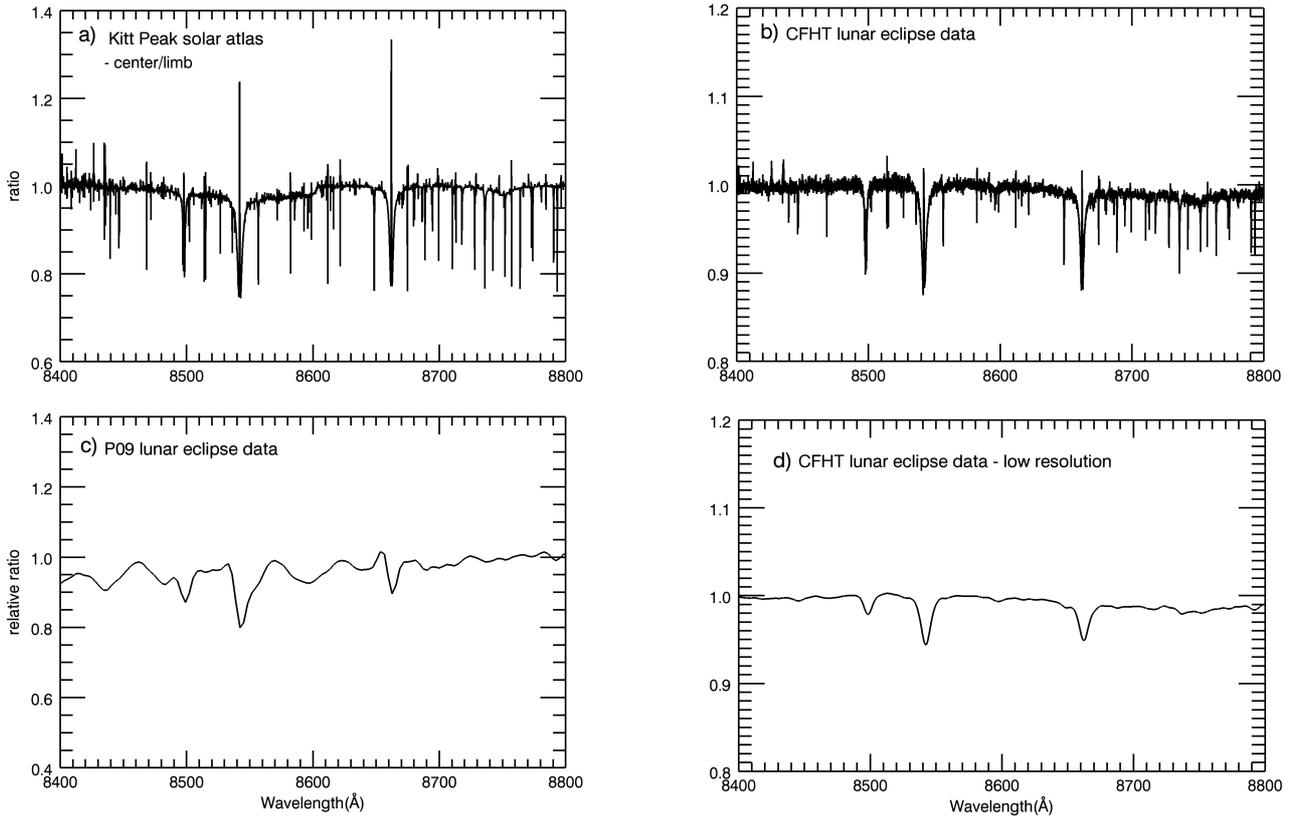}
\caption{The CLV features in the \ion{Ca}{ii} T lines. (a) Top left: ratio of the centre to limb spectrum from the Kitt Peak Solar Atlas. (b) Top right: ratio of the bright Moon spectrum to the penumbral spectrum from our CFHT data. (c) Bottom left: the \ion{Ca}{ii} T feature in P09's transmission spectrum. (d)Bottom right: the same as (b) except convolved to the resolution of P09 data.
}
         \label{Palle-CaT}
   \end{figure*}

Figure \ref{Palle-CaT} shows the detailed features at the positions of the \ion{Ca}{ii} T lines. Here the ratio of the centre to limb spectrum from the Kitt Peak Solar Atlas (Fig.~\ref{Palle-CaT}a) is used  to mimic the umbral to penumbral spectral ratio. Our CFHT data are also shown for comparison. Fig.~\ref{Palle-CaT}b is the ratio of the bright Moon spectrum to the penumbral spectrum from the CFHT observation and it shows a behaviour very similar to that constructed from  the Solar Atlas. Fig.~\ref{Palle-CaT}d shows the same data as in Fig.~\ref{Palle-CaT}b but convolved to the resolution of the P09 data.
It is clear that the CLV feature in the CFHT data is similar to the spectrum in P09 (Fig.~\ref{Palle-CaT}c).

The \ion{Na}{i} D line features are relatively weak in P09. This is interpreted as CLV features of the \ion{Na}{i} doublet being weaker than other strong Fraunhofer lines and consequentially not strongly apparent at low resolution.

\subsection{\ion{Na}{i} features in previous high spectral resolution observations}
\citet{Vidal2010} observed the 16 August 2008 lunar eclipse with high-resolution spectroscopy and they obtained the transmission spectrum by using the ratio of the penumbral to the bright Moon spectrum. In their resulting transmission spectrum, the \ion{Na}{i} D features are not well cancelled out and the zero level shift (an offset of the spectral counts) is applied to correct the broad features at \ion{Na}{i} D positions (Fig.~10 in V10). This correction results in a deeper `absorption' feature at the line centre, which the authors interpret to be the atomic sodium absorption of the Earth's atmosphere. However, we conclude that this is the CLV feature of the \ion{Na}{i} D lines and the feature at the line core originates from the differential Doppler broadening between the centre and limb solar spectra as discussed in Section 2.2.

\citet{Arnold2014} observed the 21 December 2010 lunar eclipse with two high-resolution ESO spectrographs: HARPS and UVES. They applied a similar method as V10 and obtained the high-resolution transmission spectrum from the penumbral spectrum. In the resulting transmission spectrum from HARPS, the \ion{Na}{i} D features occur (Fig.~9 in A14) and the authors use a Raman scattering model to correct the broad signature of the line wing (the Raman scattering will be discussed in the next section). This correction again results in a deeper 'absorption' feature at the line core, which is similar to the result in V10. 
They also applied a wavelength-dependent limb darkening to correct the penumbral spectrum. The limb darkening they used is a low-resolution model (several nanometres), thus it only corrects the limb darkening of the continuum while the CLV of the \ion{Na}{i} D line remains in the ratio of the penumbral to the bright Moon spectrum.
 
\begin{figure}
   \centering
   \includegraphics[width=0.5\textwidth]{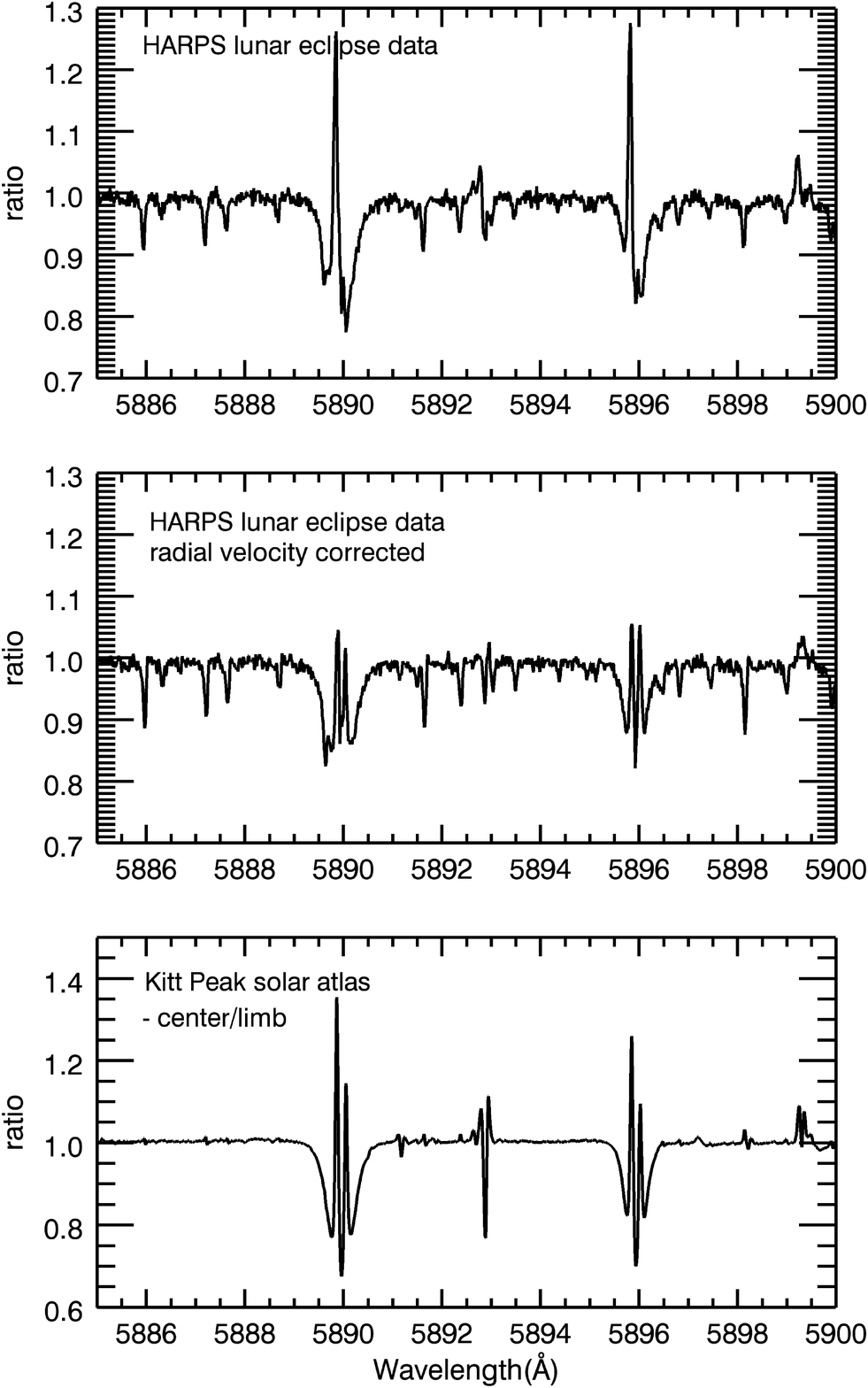}
\caption{The CLV features in the \ion{Na}{i} D lines. Top panel: ratio of the bright Moon to penumbral spectrum from the HARPS observation in December 2010. Middle panel: same as top, except with the radial velocity corrections applied. Bottom panel: the centre to limb spectrum from the Kitt Peak Solar Atlas.
}
         \label{Vidal-NaD}
   \end{figure}

We reanalysed the HARPS data from the 21 December 2010 lunar eclipse  using the ratio of the bright Moon to penumbral Moon spectrum to simulate the effective altitude calculation method (Equation 2 in A14). Fig.~\ref{Vidal-NaD} shows the result of our simulation. 
The top panel is the original ratio of the bright Moon to the penumbral Moon spectrum. This is similar to the bottom figure of Fig.~16 in A14. 
The middle panel is the ratio spectrum after the radial velocity (RV) correction. The RV difference between the bright Moon and the penumbral Moon is due to the motion between the Earth, Moon, and Sun, as well as the Rossiter-McLaughlin effect (Yan et al. in preparation).
We correct this RV difference to perfectly align the solar spectral lines.
After the RV correction, the CLV feature of the \ion{Na}{i} D lines appears clearly, and is similar to Fig.~9 in A14. 
The bottom panel in Fig.~\ref{Vidal-NaD} is the ratio of the centre to limb spectrum from the Kitt Peak Solar Atlas and it corresponds closely to the feature from the lunar eclipse observation.

In fact, the \ion{Na}{i} D feature in V10 is basically the same as that in A14 except at a lower resolution.
As the spectrum of P09 is in a much lower resolution, the \ion{Na}{i} D feature in P09 is only marginally observed. 
By comparing the \ion{Na}{i} D feature in the three papers, we can see that the appearance of the CLV feature is affected by the data analysis method, spectral resolution, and the radial velocity correction.

\section{Discussion}
\subsection{The forward-scattered sunlight and the Ring effect}
The forward-scattered sunlight from the Earth's terminator 
atmosphere also contributes to the irradiance of the eclipsed Moon 
\citep{Link1972,Vollmer2008}.
In the forward-scattered sunlight, there is a Raman scattering component 
that transfers continuum photons into a solar absorption line to make 
the scattered line shallower.
This filling-in of Fraunhofer lines is called the Ring effect 
\citep{Grainger1962} and can change the solar line shape in a lunar 
eclipse spectrum \citep{Yan2014}. This Ring effect was employed by A14 
to correct the line shape of \ion{Na}{i} D doublet.
Below we quantitatively discuss the Ring effect contribution in both the 
umbral and penumbral spectra.

\begin{figure}
   \centering
   \includegraphics[width=0.5\textwidth]{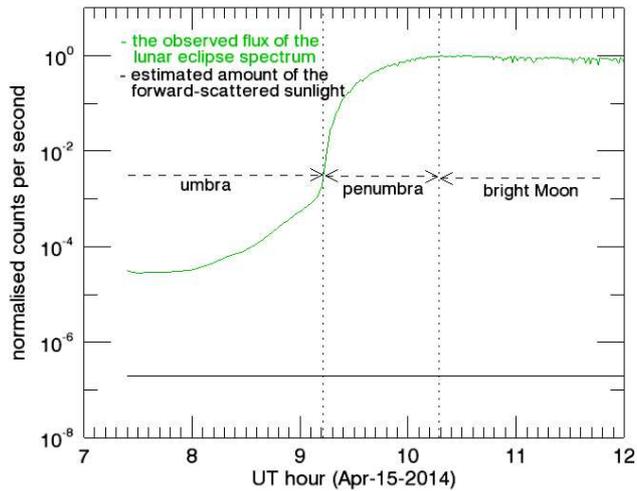}
\caption{Flux of the lunar eclipse spectrum and the forward-scattered sunlight. The green line is the total observed counts per second between 550--650 nm of the 292 lunar spectra that were observed with CFHT during the 2014-Apr-15 lunar eclipse. The numbers are normalised to a bright Moon spectrum taken at the meridian. The plot can be approximately regarded as the light curve of the eclipsed Moon, which is normalised to the flux of the unobstructed Sun (i.~e. the flux for the bright Moon). The black solid line is the estimated forward-scattered sunlight from \cite{Garcia2011} and the amount used here is $2\times 10^{-7}$ of the flux from the unobstructed Sun.
}
         \label{light-curve}
   \end{figure}

\cite{Garcia2011} modelled the forward-scattered sunlight 
spectrum during a lunar eclipse. Their result shows that, at wavelength 
around \ion{Na}{i} D doublet, the typical amount of the 
forward-scattered sunlight received at the lunar surface is $2\times 
10^{-7}$ of the flux from the unobstructed Sun (i.~e. the bright Moon). 
The amount here is for the atmospheric model containing molecular 
and aerosol scattering and $\mathrm{O_3}$ absorption, i.~e. model 
dIII in \cite{Garcia2011}.
In order to estimate the contribution of the scattered light, we use our 
CFHT observation to generate an eclipse light curve. 
Fig.~\ref{light-curve} shows the observed fluxes, which are normalised to 
a bright Moon spectrum taken at the meridian. The flux of the 
forward-scattered sunlight can be regarded as constant since it is a 
diffuse component and the scattering angle changes little during the 
eclipse.
It is clear that the scattered light proportion is larger for an 
eclipse spectrum taken at a position close to the umbral center where 
the irradiance of the lunar surface is small.
For the umbral spectrum, the scattered light proportion can range from 
0.01 to $10^{-4}$. For the penumbral spectrum, the proportion is 
determined mainly by the portion of the unobstructed solar disk and 
ranges from $10^{-4}$ to $2 \times 10^{-7}$.
For example, the amount of  scattered light is on the order of 
$10^{-5}$ for Penumbra A (Fig.~\ref{PMview}) where about 1/6 of the 
solar disk can be seen.

In the forward-scattered sunlight, the rotational Raman 
scattering transfers a few percent of the continuum photons into the 
Fraunhofer lines at UV to optical wavelengths \citep{Noxon1979, 
Langford2007}.
For an eclipse spectrum taken close to the umbral center, this 
Raman scattering transfers roughly $10^{-4}$ of the continuum into the 
lines after taking  the forward-scattering sunlight 
proportion in the total irradiance into account. The spectra taken at a position away 
from the umbral center will have a lower filling-in effect. Here the 
estimation is for wavelengths around 600 nm, however, the filling-in 
effect can be significantly stronger at the blue or UV wavelengths where 
the total umbral irradiance is lower and consequentially the 
forward-scattering proportion is larger.
For a penumbral spectrum like Penumbra A, the Raman 
scattering transfers less than $10^{-6}$ of the continuum photos, which 
is a negligible amount.

From the above estimates, we conclude that the Ring effect 
can be important in the umbra especially for the spectra taken close to 
the umbral center, and the line shape of the strong Fraunhofer line is 
affected by both the Ring effect and  CLV effect.
However, the line shape in the penumbral spectrum is dominated by 
the CLV effect, and the Ring effect can be safely ignored.

\subsection{Telluric Sodium absorption}
The sodium layer existing in the Earth's upper mesosphere originates from meteoroids \citep{Plane2003}. 
Since the sodium $\mathrm{D_2-D_1}$ doublet are resonance lines, their absorption features should be imprinted on the transmission spectrum if there is a significant sodium layer in the observed atmosphere.
Using an average Na concentration profile measured by \cite{Fussen2004} and the cross-section data in \cite{Fussen2010}, we are able to model the Na absorption for the integrated Earth atmosphere. The Na transmission spectrum for each tangential path with a minimum atmospheric altitude is calculated, and then the spectra with the minimum altitudes from 0 km to 120 km are integrated to obtain the integrated absorption spectrum.
This modelled absorption spectrum is convolved to the instrumental resolution for ESPaDOnS/CFHT and the equivalent width of the $\mathrm{D_1}$ line is calculated to be 0.007 $\,\AA$ with a maximum absorption depth of 0.08. 

\begin{figure}
   \centering
   \includegraphics[width=0.5\textwidth]{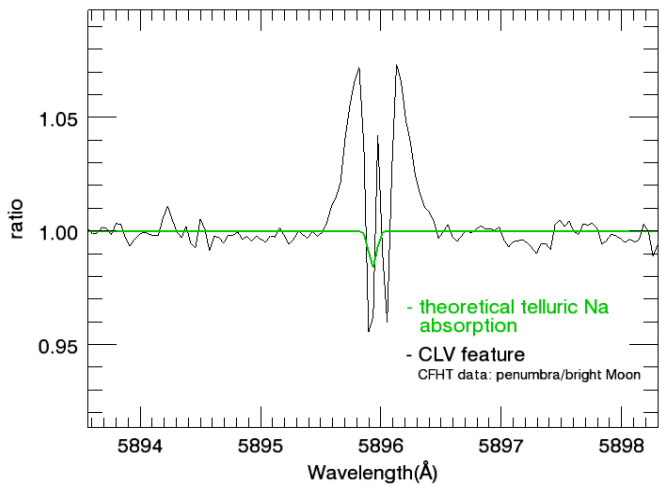}
\caption{Theoretical telluric Na absorption and the CLV feature for the \ion{Na}{i} $\mathrm{D_1}$ line (5896$\,\AA$). A RV correction has been applied to the CFHT spectra. The CLV feature still dominates the line profile in a penumbral spectrum, which makes the telluric Na absorption difficult to detect.
}
         \label{telluric-Na}
   \end{figure}

As stated before, the penumbral spectrum is predominantly the direct sunlight with a small component of light transmitted from the Earth atmosphere. The actual ratio between the direct sunlight and transmitted sunlight depends on the portion of the solar disk that is not occulted by the Earth. Fig.~\ref{telluric-Na} shows an example of the theoretical Na absorption feature in a penumbral spectrum when 1/6 of the solar disk can be seen (i.e. similar to the Penumbra A in Fig.~\ref{PMview}). The CLV feature of the corresponding penumbral to bright Moon ratio spectrum is also shown for comparison.

We emphasise that the CLV features shown here are from observations, not a theoretical model. It is difficult to theoretically model the precise line profile for a given penumbral spectrum because the line profile changes quickly with time as the eclipse proceeds (as demonstrated in Fig.~\ref{Ha-change} with H$\alpha$).
Also the Rossiter-McLaughlin effect and the convective blueshift contribute to the change of the line profile and the line position.
Thus, although the telluric sodium doublet should theoretically be present in the penumbral spectrum, its relatively weak absorption features are masked by the significant line profile change of the solar sodium doublet.

\subsection{The CLV effect in exoplanet transits}
The standard way of detecting atomic or molecular species in an exoplanet atmosphere is by observing the corresponding absorption lines in and out of transit. 
Charbonneau et al. (2002) detected \ion{Na}{i} for the first time in an exoplanet atmosphere using the transit spectro-photometric method. They observed the spectra of the \ion{Na}{i} D lines in \object{HD 209458} while the planet was transiting the star and compared this with data out of transit. The absorption depth of \ion{Na}{i} D lines in transit was seen to be deeper than that out of transit, which indicates the existence of neutral sodium somewhere in the planetary atmosphere.
Since the planet blocks different parts of the stellar disk during the transit, the CLV effect of the stellar lines needs to be considered when comparing the line depths. \cite{Charbonneau2002} modelled the effect of the CLV of the \ion{Na}{i} D lines and found the difference of the absorption depth between in-transit and out-of-transit caused by the CLV effect to be $1.5\times 10^{-5}$ for a  bandwidth of 12 $\AA$ (5887--5899 $\AA$). This value is at the level of photon-noise and its effect on the \ion{Na}{i} detection in the atmosphere of \object{HD 209458b} is negligible given that the actual observed difference of the \ion{Na}{i} D lines depth is $-23.2\times 10^{-5}$.
Thus it is safe to ignore the CLV effect for the \ion{Na}{i} detection with a broad bandwidth like 12 $\AA$ used in \cite{Charbonneau2002}.
However, as can be seen from Fig.~\ref{Na-compare}, the \ion{Na}{i} D line CLV feature is prominent in a high-resolution spectrum and a 12 $\AA$ bandwidth will smear the CLV feature. If the observation is performed at a higher resolution that can resolve  the line profile well, and a narrower bandwidth is used, the \ion{Na}{i} D CLV feature would become more prominent. 

As can be seen from the solar limb to centre ratio spectrum in Fig.~\ref{overall}b, the \ion{Na}{i} D CLV feature is relatively weak and becomes more significant for other strong Fraunhofer lines such as H$\alpha$ and \ion{Ca}{ii} H\&K.
Observations at blue and UV wavelengths should deal more carefully with the CLV since the effect becomes more prominent at shorter wavelengths.

\begin{figure}
   \centering
   \includegraphics[width=0.5\textwidth]{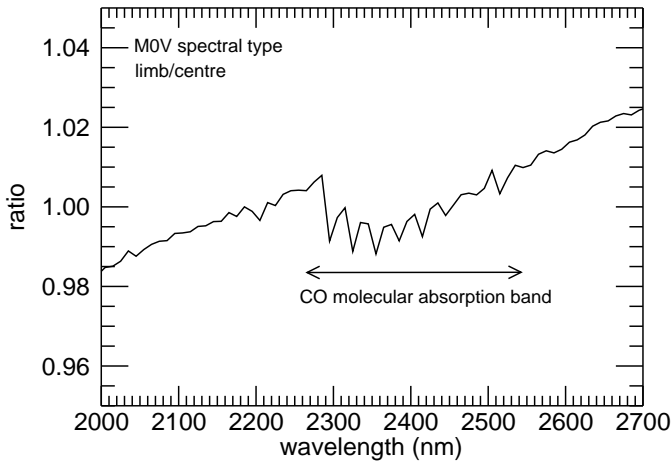}
\caption{The CLV features of a M0V type star around the CO band. The spectrum shown is the flux ratio of the limb spectrum ($\mu=0.2$) to the centre spectrum from the Kurucz stellar model. The ratio is normalised to 1 at 2200 nm.
The CO absorption band shows a different limb darkening compared to the adjacent continuum.
}
         \label{Mstar}
   \end{figure}

Late-type dwarf stars are thought to be promising targets for exoplanet atmosphere detection because the planetary transit depth is larger due to the relatively large planet-star radius ratio.
However, the CLV of the strong molecular absorption lines in these stars could affect the corresponding molecular detections in exoplanet atmospheres. Fig.~\ref{Mstar} shows the CO band CLV feature in an M-type dwarf calculated from a Kurucz grid model\footnote{http://kurucz.harvard.edu/grids.html}. 
From the model, we find that the CO band is deeper in the limb spectrum than in the centre spectrum. This indicates that when a planet transits the centre part of an M-type dwarf, the CO band in the observed spectrum will be deeper compared to the spectrum out of transit, which could potentially be interpreted as a planetary CO absorption feature.

We are currently performing a more detailed quantitative analysis of the CLV effect during exoplanet transits.
Here we mention three general aspects that affect the strength of the CLV feature.
The first concerns the star itself. The CLV effect depends on the physical parameters and the chemical abundances of the star so that different stellar types and different spectral lines will exhibit different CLV signatures. 
The second aspect is the geometry of the planet transit system, including both the planet-to-star radius ratio, which determines the obscured proportion of the stellar disk and the impact parameter of the planet's transit trajectory on the stellar disk. 
The planet will transit only the limb part of the stellar disk if $b$ is close to 1 ($b$ is the impact parameter in the unit of the stellar radius), while both the limb and the centre of the stellar disk will be blocked if $b = 0$ (i.e. edge-on).
 As can be seen from the lunar eclipse data in Fig.~\ref{Ha-change}, the actual CLV feature varies with the phase of the transit.
The third aspect is related to the observational method. For example, the spectral resolution and  the chosen bandwidth for the analysis will determine the apparent strength of the CLV effect.

Consequently the theoretical CLV modelling work for different stellar types needs to be well-chosen for the precise characterisation of exoplanet atmospheres. An additional application of these studies is the employment of the CLV effects during the exoplanet transit to serve as a probe for the study of the star itself \citep{Dravins2014}.

\section{Conclusions}
Using a sequence of lunar eclipse spectra observed with the CFHT, we have demonstrated the significant center-to-limb variations of the  solar Fraunhofer lines. Generally, the Fraunhofer lines are deeper in the centre spectrum than in the limb spectrum and the CLV features are stronger towards the blue and UV wavelength range. The CLV of the \ion{Na}{i} D lines is weaker compared to other strong lines. 

This CLV effect is especially important for studying the atom or ion absorptions in the transmission spectrum of the Earth's atmosphere. For  previous lunar eclipse observations, the \ion{Ca}{ii} absorption feature in \cite{Palle2009} and the \ion{Na}{i} D lines feature in \cite{Vidal2010} and \cite{Arnold2014} are  shown to be most likely due predominately to the CLV of the corresponding solar lines instead of the \ion{Na}{i} or \ion{Ca}{ii} absorption in the Earth's atmosphere.
Although there are variable \ion{Ca}{ii} and \ion{Na}{i} layers in the Earth's atmosphere, the reliable detection of their relatively weak absorption features is difficult because of the CLV features.

The CLV effect needs to be considered in searches for atom or ion absorptions in exoplanet atmospheres using the transit technique, especially when observing at high spectral resolution or at blue or UV wavelengths where the CLV is prominent.
Molecular absorptions in M-type stars also exhibit CLV features, thus  future molecular detections in planets transiting M-type stars should employ an accurate stellar model.

\begin{acknowledgements}
This research uses data obtained through the Telescope Access Program (TAP), which is funded by the National Astronomical Observatories, Chinese Academy of Sciences, and the Special Fund for Astronomy from the Ministry of Finance. 
The study is supported by the National Natural Science Foundation of China under grants Nos. 11390371 and 11233004.
We thank the CFHT team for the observations.
The authors thank the referee for useful suggestions.
Fei Yan acknowledges the support from ESO-NAOC studentship.
\end{acknowledgements}

\bibliographystyle{aa} 

\bibliography{CLVpaper}

\end{document}